\documentclass[preprintnumbers,amsmath,amssymbm,preprint]{revtex4}
\usepackage{epsfig}
\usepackage{graphicx}

\begin{document}
\title{The gravitational two-body problem in the vicinity of the light ring: Insights from
the black-hole-ring toy model}
\author{Shahar Hod}
\address{The Ruppin Academic Center, Emeq Hefer 40250, Israel}
\address{ }
\address{The Hadassah Institute, Jerusalem 91010, Israel}
\date{\today}

\begin{abstract}
The physical properties of an axisymmetric black-hole-ring system
are studied analytically within the framework of general relativity
to second order in the dimensionless mass ratio $\mu\equiv m/M$.
In particular, we analyze the asymptotic behaviors of the
binding-energy and the total angular-momentum of the two-body system
in the vicinity of the light ring at $R=3M$, where the circular
orbit becomes null. We find that both quantities diverge {\it
quadratically} in $\mu(1-3M/R)^{-1}$ at the light ring. The reported
divergent behavior of the physical quantities stems from the second
order spin-orbit interaction between the black hole and the orbiting
object (the dragging of inertial frames by the orbiting ring). It is
shown that this composed black-hole-ring toy model captures some of
the essential features of the conservative dynamics of the
(astrophysically more interesting) black-hole-particle system. In
particular, we show that both systems share the {\it same} quadratic
divergent behavior of the physical quantities near the light ring.
Moreover, we prove that both systems are characterized by the {\it
same} ratio ${{E^{(2)}(R\to 3M)}\over{J^{(2)}(R\to
3M)}}={{1}\over{3\sqrt{3}}}$, where $E^{(2)}$ and $J^{(2)}$ are the
divergent second order (self-interaction) expansion coefficients of
the binding-energies and the angular-momenta of the systems,
respectively.
\end{abstract}
\maketitle

\section{Introduction.}

The gravitational two-body problem has attracted much attention over
the years from both physicists and mathematicians, see
\cite{Gold,Bar,Chan,Shap,CarC,Fav,Ori,Poi,Lou1,Det1,Bar1,Det2,Sag,Kei,Sha,Dam,Bar2,Fav2,Will,Akc}
and references therein. As is well known, the two-body problem in
Newtonian gravity is characterized by a particularly elegant and
simple analytic solution \cite{Gold}. In fact, Newtonian gravity
provides a very accurate mathematical description for the dynamics
of a wide variety of astrophysical binaries in the {\it weak}-field
limit. However, the dynamics of close binaries composed of black
holes and superdense neutron stars are characterized by {\it
strong}-gravity effects which cannot be treated properly within the
limited framework of Newtonian mechanics. Instead, the dynamics of
compact binaries in the strong-gravity regime should be described by
the (mathematically more complicated) equations of the general
theory of relativity.

The Einstein equations which describe the dynamics of close binaries
in the strong-gravity regime are very complex and have no compact
analytic solution. Despite this fact, the two-body problem in
general relativity can be tackled using an approximation method
which is based on a perturbative treatment in powers of the
symmetric mass ratio \cite{Akc}
\begin{equation}\label{Eq1}
\mu\equiv {{Mm}\over {(M+m)^2}}\  ,
\end{equation}
where $M$ and $m$ are the masses of the two compact objects. (We use
gravitational units in which $G=c=1$).

Two important physical quantities which characterize the circular
two-body dynamics are the binding-energy and the total
angular-momentum of the system. The binding energy
$E_{\text{binding}}$ of the two-body system is given by the
difference between the total gravitational energy of the system at
infinity and the total gravitational energy of the system at a
finite separation:
\begin{equation}\label{Eq2}
E_{\text{binding}}(x)=E(x=0)-E(x)\  .
\end{equation}
Here
\begin{equation}\label{Eq3}
x\equiv [(M+m)\Omega]^{2/3}
\end{equation}
is a convenient invariant (and dimensionless) coordinate constructed
from the characteristic frequency $\Omega$ of the circular orbit
\cite{Akc,Notex}. The binding-energy can be expanded in powers of
the dimensionless mass ratio $\mu$:
\begin{equation}\label{Eq4}
E_{\text{binding}}(x)/M=\sum_{k=1}^{\infty}E^{(k)}(x)\cdot\mu^k\  .
\end{equation}
Likewise, the angular-momentum $J$ of the two-body system can be
expressed as a power series of the dimensionless mass ratio $\mu$:
\begin{equation}\label{Eq5}
J(x)/M^2=\sum_{k=1}^{\infty}J^{(k)}(x)\cdot\mu^k\  .
\end{equation}

In the zeroth-order ($\mu\to 0$) approximation the spacetime metric
is described by the physical parameters of the larger object (the
central black hole) while the smaller object (the `test-particle')
follows a geodesic of the black-hole spacetime. In this
test-particle limit the system is characterized by the well-known
relations \cite{Bar}
\begin{equation}\label{Eq6}
E^{(1)}(x)={{1-2x}\over{\sqrt{(1-3x)}}}-1\ \ \ \text{and}\ \ \
J^{(1)}(x)={{1}\over{\sqrt{x(1-3x)}}}\  .
\end{equation}

Realistic astrophysical binaries are often composed of an orbiting
object whose mass $m$ is smaller but {\it non}-negligible as
compared to the mass $M$ of the central black hole. In these
situations the zeroth-order (test-particle) approximation
(\ref{Eq6}) is no longer valid and one should take into account the
gravitational self-force corrections to the orbit
\cite{Ori,Poi,Lou1,Det1,Bar1,Det2,Sag,Kei,Sha,Dam,Bar2,Fav2,Will,Akc}.
These corrections (which are second-order in the symmetric mass
ratio $\mu$) take into account the {\it finite} mass of the orbiting
object.

The gravitational self-force has two distinct contributions: (1) It
is responsible for {\it dissipative} effects that cause the orbiting
particle to emit gravitational waves \cite{Ori,Fav}. (2) The
self-force due to the finite mass of the orbiting object is also
responsible for {\it conservative} corrections [of order $O(\mu^2)$]
to the binding-energy and to the total angular-momentum of the
two-body system. Following Refs.
\cite{Lou1,Det1,Bar1,Det2,Sag,Kei,Sha,Dam,Bar2,Fav2,Will,Akc}, in
the present study we shall focus on these conservative corrections
to the orbit.

\section{The black-hole-particle system near the light ring}

The conservative second-order (self-interaction) contributions to
the binding-energy and to the total angular-momentum of the two-body
(black-hole-particle) system, $E^{(2)}_{\text{particle}}(x)$ and
$J^{(2)}_{\text{particle}}(x)$, were computed most recently in
\cite{Akc,Notecon}. The mathematical expressions of these physical
quantities are quite cumbersome \cite{Akc}, but a remarkably simple
{\it quadratic divergence} of both these quantities was observed in
the vicinity of the unperturbed light ring \cite{Notelight} at
$x={1\over 3}$ \cite{Akc}:
\begin{equation}\label{Eq7}
E^{(2)}_{\text{particle}}(z\to 0)=\Big({1\over 27}-{1\over
12}\zeta\Big)\cdot z^{-2}\ \ \ \text{and}\ \ \
J^{(2)}_{\text{particle}}(z\to
0)=\Big({{1}\over{3\sqrt{3}}}-{{\sqrt{3}}\over{4}}\zeta\Big)\cdot
z^{-2}\ ,
\end{equation}
where
\begin{equation}\label{Eq8}
z\equiv 1-3x\  .
\end{equation}
Here $\zeta$ is a ``fudge" factor which was introduced in
\cite{Akc}. Using {\it numerical} computations, the value of this
fudge factor was estimated in \cite{Akc} to be
\begin{equation}\label{Eq9}
\zeta\approx 1\  .
\end{equation}

We would like to point out that the physical quantities $E^{(2)}(x)$
and $J^{(2)}(x)$ satisfy the simple relation [see Eq. (\ref{Eq7})]
\begin{equation}\label{Eq10}
{{E^{(2)}_{\text{particle}}(z\to
0)}\over{J^{(2)}_{\text{particle}}(z\to 0)}}={{1}\over{3\sqrt{3}}}\
\end{equation}
in the vicinity of the light ring. It is worth emphasizing that this
relation holds true {\it regardless} of the value of the
(numerically computed) fudge factor $\zeta$.

It is worth noting that the quadratic divergence of the physical
quantities which characterize the dynamics of the
black-hole-particle system near the light ring [see Eq. (\ref{Eq7})]
could only be inferred using {\it numerical} computations, see Fig.
2 of \cite{Akc}. The main goal of the present Letter is to analyze a
simple toy model which captures, at least qualitatively, some
important features of the fundamental two-body problem in general
relativity. In particular, we would like to provide an {\it
analytical} explanation for the {\it quadratic divergence} of the
self-interaction quantities $E^{(2)}(x)$ and $J^{(2)}(x)$ in the
vicinity of the light ring.

\section{The black-hole-ring system near the light ring}

In a recent paper \cite{Hodn} we proposed to model the conservative
behavior of the two-body system using the analytically solvable
model of a stationary axisymmetric ring of particles in orbit around
a central black hole. This composed system was analyzed in detail by
Will \cite{Will}. We have shown \cite{Hodn} that this toy model
captures, at least qualitatively, some important features of the
{\it conservative} dynamics of the (astrophysically more relevant)
black-hole-particle system \cite{Notegw}. In particular, like the
orbiting particle, the rotating ring can drag the generators of the
black-hole horizon \cite{Will}.

It is well-known \cite{Will} that local inertial frames are {\it
dragged} by an orbiting object. In fact, because of the dragging of
inertial frames by the orbiting object, one can have a
Schwarzschild-like black hole with zero angular-momentum
($J_{\text{H}}=0$) but with a {\it non}-zero angular-velocity
($\omega_{\text{H}}\neq 0$) [see Eq. (\ref{Eq12}) below].
Specifically, it was found in \cite{Will} that the angular-momentum
of a black hole which is perturbed by an orbiting ring is given by
\cite{Notexr}
\begin{equation}\label{Eq11}
J_{\text{H}}=4M^3\omega_{\text{H}}-8Mmx^3J^{(1)}(x)\ ,
\end{equation}
where the leading-order dimensionless angular momentum of the ring
,$J^{(1)}(x)$, is given by the expression (\ref{Eq6}). Thus, to
first-order in the symmetric mass ratio $\mu$ of the system [see Eq.
(\ref{Eq1})], a {\it zero} angular momentum black hole
($J_{\text{H}}=0$) in the composed black-hole-ring system is
characterized by a {\it non}-zero angular velocity
$\omega_{\text{H}}$ :
\begin{equation}\label{Eq12}
M\omega_{\text{H}}=2x^3J^{(1)}(x)\cdot\mu\  .
\end{equation}
The relation (\ref{Eq12}) [and, in particular, the fact that
$\omega_{\text{H}}(J_{\text{H}}=0)\neq 0]$ is a direct consequence
of the dragging of inertial frames by the orbiting ring
\cite{Will,Notebar}.

The binding-energy of the black-hole-ring system,
$E_{\text{binding}}(x)$, can be expanded in the form (\ref{Eq4}).
From the results presented in \cite{Will} for the composed
black-hole-ring system one finds after some algebra that the
leading-order expansion coefficient $E^{(1)}_{\text{ring}}(x)$ [the
coefficient of the $O(\mu)$ term in the expansion (\ref{Eq4})] is
given by the expression (\ref{Eq6}). In addition, one finds
\cite{Will} that the $O(\mu^2)$ contribution to the energy budget of
the black-hole-ring system in the vicinity of the light ring
($z\to0$) is dominated  by the divergent term
\begin{equation}\label{Eq13}
\mu E^{(2)}_{\text{ring}}(z\to 0)=-{4\over
{27z}}M\omega_{\text{H}}J^{(1)}\ .
\end{equation}
The expression (\ref{Eq13}) represents a spin-orbit interaction
between the spinning black hole (which is characterized by the
horizon angular velocity $\omega_{\text{H}}$) and the rotating ring
(which is characterized by the angular momentum $mJ^{(1)}$). It is
worth emphasizing that $\omega_{\text{H}}$ is linear in the mass $m$
of the orbiting ring [see Eq. (\ref{Eq12})]. Thus, this spin-orbit
interaction term represents a second-order self-interaction term of
order $O(\mu^2)$. Taking cognizance of Eqs. (\ref{Eq6}),
(\ref{Eq12}), and (\ref{Eq13}), we find \cite{Notethr}
\begin{equation}\label{Eq14}
E^{(2)}_{\text{ring}}(z\to 0)=-{{8}\over{243}}\cdot z^{-2}\
\end{equation}
in the vicinity of the light ring.

Likewise, the total angular-momentum of the black-hole-ring system,
$J(x)$, can be expanded in the form (\ref{Eq5}). From the results
presented in \cite{Will} for the composed black-hole-ring system one
finds after some algebra that the leading-order expansion
coefficient $J^{(1)}_{\text{ring}}(x)$ [the coefficient of the
$O(\mu)$ term in the expansion (\ref{Eq5})] is given by (\ref{Eq6}).
In addition, one finds \cite{Will} that the $O(\mu^2)$ contribution
to the angular-momentum of the black-hole-ring system in the
vicinity of the light ring ($z\to0$) is dominated  by the divergent
term
\begin{equation}\label{Eq15}
\mu J^{(2)}_{\text{ring}}(z\to 0)=-{4\over
3}M\omega_{\text{H}}z^{-3/2}\ .
\end{equation}
Taking cognizance of Eqs. (\ref{Eq6}), (\ref{Eq12}), and
(\ref{Eq15}), we find \cite{Notethr}
\begin{equation}\label{Eq16}
J^{(2)}_{\text{ring}}(z\to 0)=-{{8}\over{27\sqrt{3}}}\cdot z^{-2}\
\end{equation}
in the vicinity of the light ring.

It is worth emphasizing that the perturbation expansions (\ref{Eq4})
and (\ref{Eq5}) are valid in the regime $E^{(2)}(x)\cdot\mu^2\ll
E^{(1)}(x)\cdot\mu\ll 1$ [and likewise $J^{(2)}(x)\cdot\mu^2\ll
J^{(1)}(x)\cdot\mu\ll 1$]. Thus, the divergent behaviors
(\ref{Eq14}) and (\ref{Eq16}) are valid in the regime
\begin{equation}\label{Eq17}
\mu^{2/3}\ll z\ll 1\  .
\end{equation}

Inspection of Eqs. (\ref{Eq14}) and (\ref{Eq16}) reveals that the
binding-energy and the total angular-momentum of the black-hole-ring
system diverge {\it quadratically} in $\mu z^{-1}$ at the light
ring. Remarkably, the physical quantities of the original
black-hole-particle system share this same divergent behavior (that
is, a {\it quadratic} divergence in $\mu z^{-1}$) in the vicinity of
the light ring, see Eq. (\ref{Eq7}).

Moreover, the physical quantities $E^{(2)}_{\text{ring}}$ and
$J^{(2)}_{\text{ring}}$ which characterize the black-hole-ring
system satisfy the simple ratio
\begin{equation}\label{Eq18}
{{E^{(2)}_{\text{ring}}(z\to 0)}\over{J^{(2)}_{\text{ring}}(z\to
0)}}={{1}\over{3\sqrt{3}}}\  ,
\end{equation}
which is {\it identical} to the corresponding ratio
${{E^{(2)}_{\text{particle}}(z\to
0)}\over{J^{(2)}_{\text{particle}}(z\to 0)}}={{1}\over{3\sqrt{3}}}$
[see Eq. (\ref{Eq10})] satisfied by the physical quantities of the
original black-hole-particle system!

\section{Summary and discussion}

The physical properties of a black-hole-ring system were analyzed in
the vicinity of the photon orbit at $R=3M$, where the circular orbit
of the ring becomes null. We have shown that this simple toy model
may capture some important features of the conservative dynamics of
the (physically more interesting) black-hole-particle system. In
particular, this model provides a simple analytic explanation for
the {\it quadratic divergence} of the self-interaction quantities
$E^{(2)}(x)$ and $J^{(2)}(x)$ in the vicinity of the light ring, see
Eqs. (\ref{Eq14}) and (\ref{Eq16}).

Moreover, we have shown that the black-hole-particle system and the
black-hole-ring system share the {\it same} relation
\begin{equation}\label{Eq19}
{{E^{(2)}_{\text{particle}}(R\to
3M)}\over{J^{(2)}_{\text{particle}}(R\to
3M)}}={{E^{(2)}_{\text{ring}}(R\to
3M)}\over{J^{(2)}_{\text{ring}}(R\to 3M)}}={{1}\over{3\sqrt{3}}}
\end{equation}
between the second-order expansion coefficients.

The present toy model suggests that the second-order spin-orbit
interaction between the black hole and the orbiting object [the
dragging of inertial frames by the orbiting object, see Eq.
(\ref{Eq12})] is the main element determining the observed
(quadratic) divergent behavior of the self-interaction quantities in
the vicinity of the light ring.

Finally, it is worth pointing out the simple relations [see Eqs.
(\ref{Eq7}), (\ref{Eq14}), and (\ref{Eq16})]
\begin{equation}\label{Eq20}
{{E^{(2)}_{\text{particle}}(R\to
3M)}\over{E^{(2)}_{\text{ring}}(R\to
3M)}}={{J^{(2)}_{\text{particle}}(R\to
3M)}\over{J^{(2)}_{\text{ring}}(R\to 3M)}}={81\over
32}\Big(\zeta-{4\over 9}\Big)\
\end{equation}
between the self-interaction quantities of the black-hole-particle
system and the corresponding physical quantities of the
black-hole-ring system. We note that these ratios would be equal to
1 {\it if} the fudge factor $\zeta$ [see Eq. (\ref{Eq9})] equals
$68/81$.

\bigskip
\noindent
{\bf ACKNOWLEDGMENTS}
\bigskip

This research is supported by the Carmel Science Foundation. I thank
Yael Oren, Arbel M. Ongo and Ayelet B. Lata for helpful discussions.



\begin{thebibliography}{99}

\bibitem{Gold} H. Goldstein, {\it Classical Mechanics},
Addison-Wesley, Reading (1980); P. A. Sundararajan, Ph.D. Thesis,
Massachusetts Institute of Technology (2009).

\bibitem{Bar} J. M. Bardeen, W. H. Press and S. A. Teukolsky, Astrophys. J. {\bf 178}, 347 (1972).

\bibitem{Chan} S. Chandrasekhar, {\it The Mathematical Theory of Black
Holes}, (Oxford University Press, New York, 1983).

\bibitem{Shap} S. L. Shapiro and S. A. Teukolsky, {\it Black holes, white dwarfs,
and neutron stars: The physics of compact objects} (Wiley, New York,
1983).

\bibitem{CarC} V. Cardoso, A. S. Miranda, E. Berti, H. Witek and V. T.
Zanchin, Phys. Rev. D {\bf 79}, 064016 (2009).

\bibitem{Fav} M. Favata, Phys. Rev. D {\bf 83}, 024028 (2011).

\bibitem{Ori} A. Ori and K. S. Thorne, Phys. Rev. D. {\bf 62}, 124022
(2000); A. Buonanno and T. Damour, Phys. Rev. D {\bf 62}, 064015
(2000).

\bibitem{Poi} E. Poisson, Living Rev. Relativity {\bf 7}, 6 (2004).

\bibitem{Lou1} C. O. Lousto, Class. and Quant. Grav. {\bf 22}, S369
(2005).

\bibitem{Det1} S. Detweiler, in {\it Mass and Motion in General Relativity}, edited by
L. Blanchet, A. Spallicci, and B. Whiting (Springer, 2011).

\bibitem{Bar1} L. Barack, Class. and Quant. Grav. {\bf 26}, 213001
(2009).

\bibitem{Det2} S. Detweiler, Phys. Rev. D {\bf 77}, 124026 (2008).

\bibitem{Sag} N. Sago, L. Barack, and S. Detweiler, Phys. Rev. D {\bf 78}, 124024
(2008).

\bibitem{Kei} T. S. Keidl, A. G. Shah, J. L. Friedman, D. Kim, and L. R. Price,
Phys. Rev. D {\bf 82}, 124012 (2010).

\bibitem{Sha} A. Shah, T. Keidl, J. Friedman, D. Kim, and L. Price, Phys. Rev. D {\bf 83}, 064018
(2011).

\bibitem{Dam} T. Damour, Phys. Rev. D {\bf 81}, 024017 (2010).

\bibitem{Bar2} L. Barack and N. Sago, Phys. Rev. Lett. {\bf 102}, 191101 (2009);
L. Barack and N. Sago, Phys. Rev. D {\bf 81}, 084021 (2010).

\bibitem{Fav2} M. Favata, Phys. Rev. D {\bf 83}, 024027 (2011).

\bibitem{Will} C. M. Will, The astrophysical Journal {\bf 191}, 521 (1974);
C. M. Will, The astrophysical Journal {\bf 196}, 41 (1975).

\bibitem{Akc} S. Akcay, L. Barack, T. Damour, and N. Sago, Phys. Rev. D {\bf 86},
104041 (2012).

\bibitem{Notex} Note that $x$ reduces to $M/R$ in the test-particle
($m\to 0$) limit, where $R$ is the Schwarzschild radial coordinate
associated with the central black hole (the larger object).

\bibitem{Notecon} It should be emphasized that the authors of Ref. \cite{Akc} focused on
the {\it conservative} circular dynamics of the two-body system. It
is only in this non dissipative regime that the energy and angular
momentum of the system are conserved quantities.

\bibitem{Notelight} The light ring is often referred to as the null
circular geodesic of the black-hole spacetime.

\bibitem{Hodn} S. Hod, Phys. Rev. D {\bf 87}, 024036 (2013).

\bibitem{Notegw} It should be emphasized that this toy model, being axially-symmetric,
can not describe the most important characteristic of the
gravitational two-body problem: the emission of gravitational
radiation. Nevertheless, following Refs.
\cite{Lou1,Det1,Bar1,Det2,Sag,Kei,Sha,Dam,Bar2,Fav2,Will,Akc}, in
the present study we shall focus on the {\it conservative} dynamics
of the two-body system. That is, following Refs.
\cite{Lou1,Det1,Bar1,Det2,Sag,Kei,Sha,Dam,Bar2,Fav2,Will,Akc} we
shall ignore the emission of gravitational waves.


\bibitem{Notexr} Note that the results presented in \cite{Will} are expressed
in terms of the irreducible mass $M_{\text{ir}}$ of the black hole.
For a `bare' Schwarzschild black hole the irreducible mass coincides
with the mass $M$ of the black hole. For the black-hole-ring system
considered in \cite{Will} one finds $M_{\text{ir}}=M[1+O(\mu^2)]$.
Taking cognizance of Eq. (\ref{Eq6}), one finds that the $O(\mu^2)$
difference between $M_{\text{ir}}$ and $M$ does not affect the
leading-order divergent behaviors [see Eqs. (\ref{Eq14}) and
(\ref{Eq16}) below] of the $O(\mu^2)$ correction terms. In addition,
the results presented in \cite{Will} are expressed in terms of the
dimensionless ratio ${M_{\text{ir}}/R}$, where $R$ is the proper
circumferential radius of the ring. The invariant coordinate $x$
defined in (\ref{Eq3}) reduces to $M/R$ in the test-particle ($m\to
0$) limit. For finite $m$ values one has $x={M\over R}[1+O(\mu)]$.
Taking cognizance of Eq. (\ref{Eq6}), one finds that the $O(\mu)$
difference between $x$ and $M/R$ does not affect the leading-order
divergent behaviors [see Eqs. (\ref{Eq14}) and (\ref{Eq16}) below]
of the $O(\mu^2)$ correction terms. Finally, we note that the
symmetric mass ratio $\mu$ [see Eq. (\ref{Eq1})] is closely related
to the dimensionless mass ratio $q\equiv {m/M}$: $\mu=q[1+O(q)]$.
Taking cognizance of Eq. (\ref{Eq6}), one finds that the
leading-order divergent coefficients [see Eqs. (\ref{Eq14}) and
(\ref{Eq16}) below] of the $O(\mu^2)$ correction terms would also
describe the leading-order divergent coefficients of the
corresponding $O(q^2)$ corrections terms.

\bibitem{Notebar} In particular, note that a `bare' (unperturbed)
Schwarzschild black hole is characterized by the relation
$J_{\text{H}}=4M^3\omega_{\text{H}}$, which implies the simple
relation $\omega_{\text{H}}(J_{\text{H}}=0)=0$.

\bibitem{Notethr} Here we have used the fact that $x\to{1\over 3}$
in the vicinity of the light ring.

\end{thebibliography}
\end{document}